\documentclass[10pt,a4paper,onecolumn]{article}
\usepackage{marginnote}
\usepackage{graphicx}
\usepackage{xcolor}
\usepackage{authblk,etoolbox}
\usepackage{titlesec}
\usepackage{calc}
\usepackage{tikz}
\usepackage{hyperref}
\hypersetup{colorlinks,breaklinks=true,
            urlcolor=[rgb]{0.0, 0.5, 1.0},
            linkcolor=[rgb]{0.0, 0.5, 1.0}}
\usepackage{caption}
\usepackage{tcolorbox}
\usepackage{amssymb,amsmath}
\usepackage{ifxetex,ifluatex}
\usepackage{seqsplit}
\usepackage{xstring}

\usepackage{float}
\let\origfigure\figure
\let\endorigfigure\endfigure
\renewenvironment{figure}[1][2] {
    \expandafter\origfigure\expandafter[H]
} {
    \endorigfigure
}

\usepackage{fixltx2e} % provides \textsubscript
\usepackage[
  backend=biber,
]{biblatex}
\bibliography{paper.bib}

\let\textttOrig=\texttt
\def\texttt#1{\expandafter\textttOrig{\seqsplit{#1}}}
\renewcommand{\seqinsert}{\ifmmode
  \allowbreak
  \else\penalty6000\hspace{0pt plus 0.02em}\fi}

\makeatletter
\let\href@Orig=\href
\def\href@Urllike#1#2{\href@Orig{#1}{\begingroup
    \def\Url@String{#2}\Url@FormatString
    \endgroup}}
\def\href@Notdoi#1#2{\def\tempa{#1}\def\tempb{#2}%
  \ifx\tempa\tempb\relax\href@Urllike{#1}{#2}\else
  \href@Orig{#1}{#2}\fi}
\def\href#1#2{%
  \IfBeginWith{#1}{https://doi.org}%
  {\href@Urllike{#1}{#2}}{\href@Notdoi{#1}{#2}}}
\makeatother

\newlength{\cslhangindent}
\newlength{\csllabelwidth}
\newenvironment{CSLReferences}[3] % #1 hanging-ident, #2 entry spacing
 {% don't indent paragraphs
  \setlength{\parindent}{0pt}
  \ifodd #1 \everypar{\setlength{\hangindent}{\cslhangindent}}\ignorespaces\fi
  \ifnum #2 > 0
  \setlength{\parskip}{#2\baselineskip}
  \fi
 }%
 {}
\usepackage{calc}

\usepackage[top=3.5cm, bottom=3cm, right=1.5cm, left=1.0cm,
            headheight=2.2cm, reversemp, includemp, marginparwidth=4.5cm]{geometry}

\titleformat{\section}
  {\normalfont\sffamily\Large\bfseries}
  {}{0pt}{}
\titleformat{\subsection}
  {\normalfont\sffamily\large\bfseries}
  {}{0pt}{}
\titleformat{\subsubsection}
  {\normalfont\sffamily\bfseries}
  {}{0pt}{}
\titleformat*{\paragraph}
  {\sffamily\normalsize}

\usepackage{fancyhdr}
\makeatletter
\let\ps@plain\ps@fancy
\fancyheadoffset[L]{4.5cm}
\fancyfootoffset[L]{4.5cm}

\definecolor{linky}{rgb}{0.0, 0.5, 1.0}

\newtcolorbox{repobox}
   {colback=red, colframe=red!75!black,
     boxrule=0.5pt, arc=2pt, left=6pt, right=6pt, top=3pt, bottom=3pt}

\newcommand{\ExternalLink}{%
   \tikz[x=1.2ex, y=1.2ex, baseline=-0.05ex]{%
       \begin{scope}[x=1ex, y=1ex]
           \clip (-0.1,-0.1)
               --++ (-0, 1.2)
               --++ (0.6, 0)
               --++ (0, -0.6)
               --++ (0.6, 0)
               --++ (0, -1);
           \path[draw,
               line width = 0.5,
               rounded corners=0.5]
               (0,0) rectangle (1,1);
       \end{scope}
       \path[draw, line width = 0.5] (0.5, 0.5)
           -- (1, 1);
       \path[draw, line width = 0.5] (0.6, 1)
           -- (1, 1) -- (1, 0.6);
       }
   }

\patchcmd{\@maketitle}{center}{flushleft}{}{}
\patchcmd{\@maketitle}{center}{flushleft}{}{}
\patchcmd{\@maketitle}{\LARGE}{\LARGE\sffamily}{}{}
\def\maketitle{{%
  
  \AB@maketitle}}
\makeatletter
\renewcommand\AB@affilsepx{ \protect\Affilfont}
\renewcommand\AB@affilnote[1]{{\bfseries #1}\hspace{3pt}}
\renewcommand{\affil}[2][]%
   {\newaffiltrue\let\AB@blk@and\AB@pand
      \if\relax#1\relax\def\AB@note{\AB@thenote}\else\def\AB@note{#1}%
        \setcounter{Maxaffil}{0}\fi
        \begingroup
        \let\href=\href@Orig
        \let\texttt=\textttOrig
        \let\protect\@unexpandable@protect
        \def\thanks{\protect\thanks}\def\footnote{\protect\footnote}%
        \@temptokena=\expandafter{\AB@authors}%
        {\def\\{\protect\\\protect\Affilfont}\xdef\AB@temp{#2}}%
         \xdef\AB@authors{\the\@temptokena\AB@las\AB@au@str
         \protect\\[\affilsep]\protect\Affilfont\AB@temp}%
         \gdef\AB@las{}\gdef\AB@au@str{}%
        {\def\\{, \ignorespaces}\xdef\AB@temp{#2}}%
        \@temptokena=\expandafter{\AB@affillist}%
        \xdef\AB@affillist{\the\@temptokena \AB@affilsep
          \AB@affilnote{\AB@note}\protect\Affilfont\AB@temp}%
      \endgroup
       \let\AB@affilsep\AB@affilsepx
}
\makeatother

\renewcommand\Affilfont{\sffamily\small\mdseries}
\ifnum 0\ifxetex 1\fi\ifluatex 1\fi=0 % if pdftex
  \usepackage[T1]{fontenc}
  \usepackage[utf8]{inputenc}

\else % if luatex or xelatex
  \ifxetex
    \usepackage{mathspec}
    \usepackage{fontspec}

  \else
    \usepackage{fontspec}
  \fi
  \defaultfontfeatures{Ligatures=TeX,Scale=MatchLowercase}

\fi
\IfFileExists{upquote.sty}{\usepackage{upquote}}{}
\IfFileExists{microtype.sty}{%
\usepackage{microtype}
\UseMicrotypeSet[protrusion]{basicmath} % disable protrusion for tt fonts
}{}

\usepackage{hyperref}
\hypersetup{unicode=true,
            pdftitle={HostPhot: global and local photometry of galaxies hosting supernovae or other transients},
            pdfborder={0 0 0},
            breaklinks=true}
\let\addcontentslineOrig=\addcontentsline
\def\addcontentsline#1#2#3{\bgroup
  \let\texttt=\textttOrig\addcontentslineOrig{#1}{#2}{#3}\egroup}
\let\markbothOrig\markboth
\def\markboth#1#2{\bgroup
  \let\texttt=\textttOrig\markbothOrig{#1}{#2}\egroup}
\let\markrightOrig\markright
\def\markright#1{\bgroup
  \let\texttt=\textttOrig\markrightOrig{#1}\egroup}

\IfFileExists{parskip.sty}{%
\usepackage{parskip}
}{% else
\setlength{\parindent}{0pt}
\setlength{\parskip}{6pt plus 2pt minus 1pt}
}
\ifx\paragraph\undefined\else
\let\oldparagraph\paragraph
\renewcommand{\paragraph}[1]{\oldparagraph{#1}\mbox{}}
\fi
\ifx\subparagraph\undefined\else
\let\oldsubparagraph\subparagraph
\renewcommand{\subparagraph}[1]{\oldsubparagraph{#1}\mbox{}}
\fi

\title{HostPhot: global and local photometry of galaxies hosting
supernovae or other transients}

        \author[1, 2]{Tomás E. Müller-Bravo}
          \author[1, 2]{Lluís Galbany}
    
      \affil[1]{Institute of Space Sciences (ICE, CSIC), Campus UAB,
Carrer de Can Magrans, s/n, E-08193 Barcelona, Spain}
      \affil[2]{Institut d'Estudis Espacials de Catalunya (IEEC),
E-08034 Barcelona, Spain}
  \date{\vspace{-7ex}}

\begin{document}
\maketitle

\marginpar{

  \begin{flushleft}
  %\hrule
  \sffamily\small

  {\bfseries DOI:} \href{https://doi.org/DOI unavailable}{\color{linky}{DOI unavailable}}

  \vspace{2mm}

  {\bfseries Software}
  \begin{itemize}
    \setlength\itemsep{0em}
    \item \href{N/A}{\color{linky}{Review}} \ExternalLink
    \item \href{NO_REPOSITORY}{\color{linky}{Repository}} \ExternalLink
    \item \href{DOI unavailable}{\color{linky}{Archive}} \ExternalLink
  \end{itemize}

  \vspace{2mm}

  \par\noindent\hrulefill\par

  \vspace{2mm}

  {\bfseries Editor:} \href{https://example.com}{Pending
Editor} \ExternalLink \\
  \vspace{1mm}
    {\bfseries Reviewers:}
  \begin{itemize}
  \setlength\itemsep{0em}
    \item \href{https://github.com/Pending Reviewers}{@Pending
Reviewers}
    \end{itemize}
    \vspace{2mm}

  {\bfseries Submitted:} N/A\\
  {\bfseries Published:} N/A

  \vspace{2mm}
  {\bfseries License}\\
  Authors of papers retain copyright and release the work under a Creative Commons Attribution 4.0 International License (\href{http://creativecommons.org/licenses/by/4.0/}{\color{linky}{CC BY 4.0}}).

  \end{flushleft}
}

\hypertarget{summary}{%
\section{Summary}\label{summary}}

Type Ia supernovae (SNe Ia) have assumed a fundamental role as
cosmological distance indicators since the discovery of the accelerating
expansion rate of the universe (Perlmutter et al., 1999; Riess et al.,
1998). Correlations between their optical peak luminosity, the decline
rate of their light curves and their optical colours allow them to be
standardised, reducing their observed r.m.s scatter (e.g. Phillips,
1993; Tripp, 1998). Over a decade ago, the optical peak luminosity of
SNe Ia was found to correlate with host galaxy stellar mass, further
improving their standardisation (Kelly \& others, 2010; Lampeitl et al.,
2010; Sullivan et al., 2010). Since then, host galaxy properties have
been used in cosmological analyses of SNe Ia (Betoule et al., 2014;
Brout et al., 2019; Scolnic et al., 2018) and tremendous effort has gone
into findig the property, such as star formation rate (Rigault et al.,
2013), that fundamentally drives the correlation between SNe Ia and
their host galaxies. Furthermore, it has been noted that the local
environment in which the progenitors of SNe Ia evolve is much better at
reducing the scatter in estimated distances than the global environment,
i.e., the whole galaxy (Roman et al., 2018; Kelsey et al., 2021).
Therefore, the study of the effect of environment on SNe Ia is an active
field of research and key in future cosmological analyses.

\hypertarget{statement-of-need}{%
\section{Statement of need}\label{statement-of-need}}

\texttt{HostPhot} is an open-source Python package for measuring galaxy
photmetry, both locally and globally. Galaxy photometry is fundamental
as it is commmonly used to estimate the galaxy parameters, such as
stellar mass and star formation rate. However, the codes used to
calculate photometry by different groups can vary and there is no
dedicated package for this. The API for \texttt{HostPhot} allows the
user to extract public image cutouts of surveys, such as the Panoramic
Survey Telescope and Rapid Response System (Pan-STARRS) Data Release 1
(PS1), Dark Energy Survey (DES) and Sloan Digital Sky Survey (SDSS).
Different sets of filters are available depending on the chosen survey:
\emph{grizy} for PS1, \emph{grizY} for DES and \emph{ugriz} for SDSS.
All photometry is corrected for Milky Way dust extinction. Furthermore,
\texttt{HostPhot} also works with private data obtained by the user and
can even be easily modified to include other surveys.

The major novelty of \texttt{HostPhot} is dealing with low-redshift
galaxies (z \(<\) 0.1) as obtaining photometry of these is not as simple
as those at higher redshift. Foreground stars can be in the line of
sight of nearby galaxies, making the extraction of the photometry a
complex procedure. In addition, low-redshift galaxies have visible
structures, while at high redshift they just look like simple ellipses.
\texttt{HostPhot} is able to detect sources in the images, cross-match
them with catalogs of stars (e.g., Gaia (Gaia Collaboration, 2016)) and
remove them by applying a convolution with a 2D Gaussian kernel. This
process ensures that only stars (and in some cases other galaxies that
are not of interest) are removed, keeping the structure of the galaxy
intact.

\texttt{HostPhot} can calculate the photometry of an entire galaxy
(global) or in a given circular aperture (local) and it heavily relies
on the \texttt{Astropy} (Astropy Collaboration et al., 2018, 2013) and
\texttt{Photutils} (Bradley et al., 2021) packages for this. Local
photometry can be calculated for different circular apertures in
physical units (e.g., 4 kpc) at the redshift of the given object, as has been done on previous works (e.g., Roman et al., 2018; Kelsey et al., 2021). In
addition, as the physical size depends on the assumed cosmology, the
cosmological model can be changed by the user, suiting their needs. On
the other hand, for the global photometry, the user can choose between
using a different aperture for each filter/image or a common aperture
for all the filters/images (as done in, e.g., Wiseman et al., 2020). For the latter, \texttt{HostPhot} coadds
images in the desired filters, as selected by the user (e.g.,
\emph{riz}), and estimates the common aperture parameters from the coadd
image. The aperture used for the global photometry can also be
optimised, by increasing the size until the change in flux is
negligible, encompassing the entire galaxy. In a few cases, nearby
galaxies can have very complex structures. \texttt{HostPhot} offers the
option of interactively setting the aperture via an intuitive GUI. This
option also allows the user to test how the change in aperture shape can
affect the calculated photometry.

\texttt{HostPhot} is user-friendly and well documented\footnote{https://hostphot.readthedocs.io/en/latest/},
which allows the community to easily contribute to this package.
\texttt{HostPhot} is already being used by different groups, such as
HostFlows\footnote{https://hostflows.github.io/} and DES, and will allow
the supernova community to find exciting new scientific discoveries with
future cosmological analyses. Finally, although \texttt{HostPhot} is
mainly aimed at supernova science, it can be used in other fields in
astronomy as well.

Apart from \texttt{Astropy} and \texttt{Photutils} (Bradley et al.,
2021), \texttt{HostPhot} also relies on \texttt{sep} (Barbary, 2016b)
for global photometry, \texttt{Astroquery} (Ginsburg et al., 2019) for
image downloading and cross-matching with catalogs, reproject\footnote{https://pypi.org/project/reproject/}
for the coadds, \texttt{extinction} (Barbary, 2016a) and
sfdmap\footnote{https://github.com/kbarbary/sfdmap} for extinction
correction. Finally, \texttt{HostPhot} makes use of the following
packages as well: numpy (Harris et al., 2020), matplotlib (Hunter,
2007), pandas (McKinney \& others, 2010), pyvo (Graham et al., 2014),
ipywidgets\footnote{https://github.com/jupyter-widgets/ipywidgets} and
ipympl\footnote{https://github.com/matplotlib/ipympl}.

\hypertarget{acknowledgements}{%
\section{Acknowledgements}\label{acknowledgements}}

TEMB and LG acknowledge financial support from the Spanish Ministerio de
Ciencia e Innovación (MCIN), the Agencia Estatal de Investigación (AEI)
10.13039/501100011033 under the PID2020-115253GA-I00 HOSTFLOWS project,
and from Centro Superior de Investigaciones Científicas (CSIC) under the
PIE project 20215AT016, and the I-LINK 2021 LINKA20409. TEMB and LG are
also partially supported by the program Unidad de Excelencia Maríia de
Maeztu CEX2020-001058-M. LG also acknowledges MCIN, AEI and the European
Social Fund (ESF) ``Investing in your future'' under the 2019 Ramón y
Cajal program RYC2019-027683-I.

\hypertarget{references}{%
\section*{References}\label{references}}
\addcontentsline{toc}{section}{References}

\hypertarget{refs}{}
\begin{CSLReferences}{1}{0}
\leavevmode\hypertarget{ref-astropy2}{}%
Astropy Collaboration, Price-Whelan, A. M., Sipőcz, B. M., Günther, H.
M., Lim, P. L., Crawford, S. M., Conseil, S., Shupe, D. L., Craig, M.
W., Dencheva, N., Ginsburg, A., Vand erPlas, J. T., Bradley, L. D.,
Pérez-Suárez, D., de Val-Borro, M., Aldcroft, T. L., Cruz, K. L.,
Robitaille, T. P., Tollerud, E. J., \ldots{} Astropy Contributors.
(2018). {The Astropy Project: Building an Open-science Project and
Status of the v2.0 Core Package}. \emph{The Astronomical Journal},
\emph{156}(3), 123. \url{https://doi.org/10.3847/1538-3881/aabc4f}

\leavevmode\hypertarget{ref-astropy}{}%
Astropy Collaboration, Robitaille, T. P., Tollerud, E. J., Greenfield,
P., Droettboom, M., Bray, E., Aldcroft, T., Davis, M., Ginsburg, A.,
Price-Whelan, A. M., Kerzendorf, W. E., Conley, A., Crighton, N.,
Barbary, K., Muna, D., Ferguson, H., Grollier, F., Parikh, M. M., Nair,
P. H., \ldots{} Streicher, O. (2013). {Astropy: A community Python
package for astronomy}. \emph{Astronomy and Astrophysics}, \emph{558},
A33. \url{https://doi.org/10.1051/0004-6361/201322068}

\leavevmode\hypertarget{ref-extinction}{}%
Barbary, K. (2016a). \emph{Extinction v0.3.0}. Zenodo.
\url{https://doi.org/10.5281/zenodo.804967}

\leavevmode\hypertarget{ref-sep}{}%
Barbary, K. (2016b). SEP: Source extractor as a library. \emph{Journal
of Open Source Software}, \emph{1}(6), 58.
\url{https://doi.org/10.21105/joss.00058}

\leavevmode\hypertarget{ref-Betoule2014}{}%
Betoule, M., Kessler, R., Guy, J., Mosher, J., Hardin, D., Biswas, R.,
Astier, P., El-Hage, P., Konig, M., Kuhlmann, S., Marriner, J., Pain,
R., Regnault, N., Balland, C., Bassett, B. A., Brown, P. J., Campbell,
H., Carlberg, R. G., Cellier-Holzem, F., \ldots{} Wheeler, C. J. (2014).
{Improved cosmological constraints from a joint analysis of the SDSS-II
and SNLS supernova samples}. \emph{Astronomy and Astrophysics},
\emph{568}, A22. \url{https://doi.org/10.1051/0004-6361/201423413}

\leavevmode\hypertarget{ref-photutils}{}%
Bradley, L., Sipőcz, B., Robitaille, T., Tollerud, E., Vinícius, Z.,
Deil, C., Barbary, K., Wilson, T. J., Busko, I., Donath, A., Günther, H.
M., Cara, M., krachyon, Conseil, S., Bostroem, A., Droettboom, M., Bray,
E. M., Lim, P. L., Bratholm, L. A., \ldots{} Souchereau, H. (2021).
\emph{Astropy/photutils: 1.3.0} (Version 1.3.0) {[}Computer software{]}.
Zenodo. \url{https://doi.org/10.5281/zenodo.5796924}

\leavevmode\hypertarget{ref-Brout2019}{}%
Brout, D., Sako, M., Scolnic, D., Kessler, R., D'Andrea, C. B., Davis,
T. M., Hinton, S. R., Kim, A. G., Lasker, J., Macaulay, E., Möller, A.,
Nichol, R. C., Smith, M., Sullivan, M., Wolf, R. C., Allam, S., Bassett,
B. A., Brown, P., Castander, F. J., \ldots{} DES COLLABORATION. (2019).
{First Cosmology Results Using Type Ia Supernovae from the Dark Energy
Survey: Photometric Pipeline and Light-curve Data Release}. \emph{The
Astrophysical Journal}, \emph{874}(1), 106.
\url{https://doi.org/10.3847/1538-4357/ab06c1}

\leavevmode\hypertarget{ref-gaia}{}%
Gaia Collaboration. (2016). {The Gaia mission}. \emph{Astronomy and
Astrophysics}, \emph{595}.
\url{https://doi.org/10.1051/0004-6361/201629272}

\leavevmode\hypertarget{ref-astroquery}{}%
Ginsburg, A., Sipőcz, B. M., Brasseur, C. E., Cowperthwaite, P. S.,
Craig, M. W., Deil, C., Guillochon, J., Guzman, G., Liedtke, S., Lian
Lim, P., Lockhart, K. E., Mommert, M., Morris, B. M., Norman, H.,
Parikh, M., Persson, M. V., Robitaille, T. P., Segovia, J.-C., Singer,
L. P., \ldots{} a subset of the astropy collaboration. (2019).
{astroquery: An Astronomical Web-querying Package in Python}. \emph{The
Astronomical Journal}, \emph{157}, 98.
\url{https://doi.org/10.3847/1538-3881/aafc33}

\leavevmode\hypertarget{ref-pyvo}{}%
Graham, M., Plante, R., Tody, D., \& Fitzpatrick, M. (2014).
\emph{{PyVO: Python access to the Virtual Observatory}} (p.
ascl:1402.004). Astrophysics Source Code Library, record ascl:1402.004.

\leavevmode\hypertarget{ref-numpy}{}%
Harris, C. R., Millman, K. J., Walt, S. J. van der, Gommers, R.,
Virtanen, P., Cournapeau, D., Wieser, E., Taylor, J., Berg, S., Smith,
N. J., Kern, R., Picus, M., Hoyer, S., Kerkwijk, M. H. van, Brett, M.,
Haldane, A., Río, J. F. del, Wiebe, M., Peterson, P., \ldots{} Oliphant,
T. E. (2020). Array programming with {NumPy}. \emph{Nature},
\emph{585}(7825), 357--362.
\url{https://doi.org/10.1038/s41586-020-2649-2}

\leavevmode\hypertarget{ref-matplotlib}{}%
Hunter, J. D. (2007). Matplotlib: A 2D graphics environment.
\emph{Computing in Science Engineering}, \emph{9}(3), 90--95.

\leavevmode\hypertarget{ref-Kelly2010}{}%
Kelly, P. L., \& others. (2010). {Hubble Residuals of Nearby Type Ia
Supernovae are Correlated with Host Galaxy Masses}. \emph{The
Astrophysical Journal}, \emph{715}, 743--756.
\url{https://doi.org/10.1088/0004-637X/715/2/743}

\leavevmode\hypertarget{ref-Kelsey2021}{}%
Kelsey, L., Sullivan, M., Smith, M., Wiseman, P., Brout, D., Davis, T.
M., Frohmaier, C., Galbany, L., Grayling, M., Gutiérrez, C. P., Hinton,
S. R., Kessler, R., Lidman, C., Möller, A., Sako, M., Scolnic, D.,
Uddin, S. A., Vincenzi, M., Abbott, T. M. C., \ldots{} DES
Collaboration. (2021). {The effect of environment on Type Ia supernovae
in the Dark Energy Survey three-year cosmological sample}. \emph{Monthly
Notices of the Royal Astronomical Society}, \emph{501}(4), 4861--4876.
\url{https://doi.org/10.1093/mnras/staa3924}

\leavevmode\hypertarget{ref-Lampeitl2010}{}%
Lampeitl, H., Nichol, R. C., Seo, H.-J., Giannantonio, T., Shapiro, C.,
Bassett, B., Percival, W. J., Davis, T. M., Dilday, B., Frieman, J.,
Garnavich, P., Sako, M., Smith, M., Sollerman, J., Becker, A. C.,
Cinabro, D., Filippenko, A. V., Foley, R. J., Hogan, C. J., \ldots{}
Zheng, C. (2010). {First-year Sloan Digital Sky Survey-II supernova
results: consistency and constraints with other intermediate-redshift
data sets}. \emph{Monthly Notices of the Royal Astronomical Society},
\emph{401}(4), 2331--2342.
\url{https://doi.org/10.1111/j.1365-2966.2009.15851.x}

\leavevmode\hypertarget{ref-pandas}{}%
McKinney, W., \& others. (2010). Data structures for statistical
computing in python. \emph{Proceedings of the 9th Python in Science
Conference}, \emph{445}, 51--56.

\leavevmode\hypertarget{ref-Perlmutter1999}{}%
Perlmutter, S., Aldering, G., Goldhaber, G., Knop, R. A., Nugent, P.,
Castro, P. G., Deustua, S., Fabbro, S., Goobar, A., Groom, D. E., Hook,
I. M., Kim, A. G., Kim, M. Y., Lee, J. C., Nunes, N. J., Pain, R.,
Pennypacker, C. R., Quimby, R., Lidman, C., \ldots{} Project, T. S. C.
(1999). {Measurements of {\(\Omega\)} and {\(\Lambda\)} from 42
High-Redshift Supernovae}. \emph{The Astrophysical Journal},
\emph{517}(2), 565--586. \url{https://doi.org/10.1086/307221}

\leavevmode\hypertarget{ref-Phillips1993}{}%
Phillips, M. M. (1993). {The Absolute Magnitudes of Type IA Supernovae}.
\emph{Astrophysical Journal Letters}, \emph{413}, L105.
\url{https://doi.org/10.1086/186970}

\leavevmode\hypertarget{ref-Riess1998}{}%
Riess, A. G., Filippenko, A. V., Challis, P., Clocchiatti, A., Diercks,
A., Garnavich, P. M., Gilliland, R. L., Hogan, C. J., Jha, S., Kirshner,
R. P., Leibundgut, B., Phillips, M. M., Reiss, D., Schmidt, B. P.,
Schommer, R. A., Smith, R. C., Spyromilio, J., Stubbs, C., Suntzeff, N.
B., \& Tonry, J. (1998). {Observational Evidence from Supernovae for an
Accelerating Universe and a Cosmological Constant}. \emph{The
Astronomical Journal}, \emph{116}(3), 1009--1038.
\url{https://doi.org/10.1086/300499}

\leavevmode\hypertarget{ref-Rigault2013}{}%
Rigault, M., Copin, Y., Aldering, G., Antilogus, P., Aragon, C., Bailey,
S., Baltay, C., Bongard, S., Buton, C., Canto, A., Cellier-Holzem, F.,
Childress, M., Chotard, N., Fakhouri, H. K., Feindt, U., Fleury, M.,
Gangler, E., Greskovic, P., Guy, J., \ldots{} Weaver, B. A. (2013).
{Evidence of environmental dependencies of Type Ia supernovae from the
Nearby Supernova Factory indicated by local H{\(\alpha\)}}.
\emph{Astronomy and Astrophysics}, \emph{560}, A66.
\url{https://doi.org/10.1051/0004-6361/201322104}

\leavevmode\hypertarget{ref-Roman2018}{}%
Roman, M., Hardin, D., Betoule, M., Astier, P., Balland, C., Ellis, R.
S., Fabbro, S., Guy, J., Hook, I., Howell, D. A., Lidman, C., Mitra, A.,
Möller, A., Mourão, A. M., Neveu, J., Palanque-Delabrouille, N.,
Pritchet, C. J., Regnault, N., Ruhlmann-Kleider, V., \ldots{} Sullivan,
M. (2018). {Dependence of Type Ia supernova luminosities on their local
environment}. \emph{Astronomy and Astrophysics}, \emph{615}, A68.
\url{https://doi.org/10.1051/0004-6361/201731425}

\leavevmode\hypertarget{ref-Scolnic2018}{}%
Scolnic, D. M., Jones, D. O., Rest, A., Pan, Y. C., Chornock, R., Foley,
R. J., Huber, M. E., Kessler, R., Narayan, G., Riess, A. G., Rodney, S.,
Berger, E., Brout, D. J., Challis, P. J., Drout, M., Finkbeiner, D.,
Lunnan, R., Kirshner, R. P., Sand ers, N. E., \ldots{} Smith, K. W.
(2018). {The Complete Light-curve Sample of Spectroscopically Confirmed
SNe Ia from Pan-STARRS1 and Cosmological Constraints from the Combined
Pantheon Sample}. \emph{The Astrophysical Journal}, \emph{859}(2), 101.
\url{https://doi.org/10.3847/1538-4357/aab9bb}

\leavevmode\hypertarget{ref-Sullivan2010}{}%
Sullivan, M., Conley, A., Howell, D. A., Neill, J. D., Astier, P.,
Balland, C., Basa, S., Carlberg, R. G., Fouchez, D., Guy, J., Hardin,
D., Hook, I. M., Pain, R., Palanque-Delabrouille, N., Perrett, K. M.,
Pritchet, C. J., Regnault, N., Rich, J., Ruhlmann-Kleider, V., \ldots{}
Walker, E. S. (2010). {The dependence of Type Ia Supernovae luminosities
on their host galaxies}. \emph{Monthly Notices of the Royal Astronomical
Society}, \emph{406}(2), 782--802.
\url{https://doi.org/10.1111/j.1365-2966.2010.16731.x}

\leavevmode\hypertarget{ref-Tripp1998}{}%
Tripp, R. (1998). {A two-parameter luminosity correction for Type IA
supernovae}. \emph{Astronomy and Astrophysics}, \emph{331}, 815--820.

\leavevmode\hypertarget{ref-Wiseman2020}{}%
Wiseman, P., Smith, M., Childress, M., Kelsey, L., M{\"o}ller, A., Gupta, R.~R. \ldots{} DES Collaboration (2020). {Supernova host galaxies in the dark energy survey: I. Deep coadds, photometry, and stellar masses}. \emph{Monthly
Notices of the Royal Astronomical Society}, \emph{495}, 4.
\url{https://doi.org/10.1093/mnras/staa1302}

\end{CSLReferences}

\end{document}